\documentclass[twocolumn]{aastex701}
\usepackage{amsmath}
\newcommand{\di}{d_{\mathrm{i}}}
\newcommand{\vA}{v_{\mathrm{A}}}
\newcommand{\Dfour}{D_{\mathrm{4sc}}}
\newcommand{\Done}{D_{\mathrm{1sc}}}
\shorttitle{Reconnection Exhaust Thickness}
\shortauthors{Chettri et al.}

\begin{document}

\title{Direct Four-Spacecraft Measurement of Reconnection Exhaust Thickness in the Solar Wind}

\correspondingauthor{Hemam D. Singh}
\email{hemam.singh@nsut.ac.in}

\author[0009-0000-1368-9263]{Mani K Chettri}
\affiliation{Department of Physics, Sikkim University, Gangtok 737102, India}
\email{mkchettri.22pdpy03@cus.ac.in}

\author[0000-0003-3955-7116]{Rupak Mukherjee}
\affiliation{Department of Physics, Sikkim University, Gangtok 737102, India}
\email{rmukherjee@cus.ac.in}

\author[0009-0003-3061-8944]{Hemam D. Singh}
\affiliation{Department of Physics, Netaji Subhas University of Technology, New Delhi 110078, India}
\email{hemam.singh@nsut.ac.in}

\begin{abstract}

The thickness of a reconnecting solar wind current sheet is normally inferred from a single spacecraft as the
product of the boundary-normal speed and the crossing duration. Multipoint measurements have tested this
method for generic solar wind current sheets and constrained the geometry of reconnection exhausts, but a
direct comparison with single-spacecraft thickness estimates for the same reconnection-associated crossings
is still lacking. We analyze four current-sheet crossings observed by Magnetospheric Multiscale in the
pristine solar wind on 2017 November 10---the two boundaries of a confirmed reconnection exhaust and two
nearby current sheets. The tetrahedron separation was $\approx16$~km. Cross-correlation of the magnetic-field
ramps gives millisecond-level lag uncertainties, boundary normals to $4^{\circ}$--$8^{\circ}$, and speeds to
5\%--8\%, while single-threshold timing at these separations is noise dominated. The exhaust edges are
$29$--$31$ ion inertial lengths ($\di$) thick. The two edge normals differ by $22^{\circ}$ (68\% interval
$[19^{\circ},26^{\circ}]$), resolving a nonparallel exhaust geometry. At the exhaust edges, the measured
boundary speeds agree with three standard single-spacecraft estimates to 4\%--15\% and remain within about
one Alfv\'en speed of the plasma motion. Estimating the normal from one spacecraft is the larger source of
error, though the thickness is still recovered to within tens of percent when a degenerate minimum-variance
solution is rejected. At the actively reconnecting sheet, the boundary speed exceeds the local plasma speed
along the normal by $29\%\pm10\%$.

\end{abstract}

\keywords{Solar wind (1534) --- Interplanetary discontinuities (820) --- Space plasmas (1544) --- Plasma physics (2089)}

\section{Introduction}\label{sec:intro}

Petschek-type reconnection exhausts \citep{Petschek1964} are common in the solar wind at current sheets of all
shear angles \citep{Gosling2005,Phan2006,Davis2006,Gosling2012,GoslingPhan2013,Mistry2017}. A basic property
of these layers, their thickness, enters estimates of reconnection-driven energy conversion, of the exhaust
opening angle and distance to the X line, and of the role of reconnection in dissipating solar wind turbulence.
Almost all such measurements in the solar wind are made with a single spacecraft, from the early discontinuity
surveys to recent Parker Solar Probe and near-Mars results \citep{Gosling2005,Phan2020,Zhang2026}. With one
spacecraft the thickness must be inferred as the product of a boundary speed and the crossing duration,
$D=|V\!\cdot\!\hat{N}|\,\delta t$. For reconnection-associated boundaries, however, timing-derived propagation
and thickness have not been compared directly with standard single-spacecraft estimates for the same crossings.
The normal $\hat{N}$ is usually taken from minimum variance analysis (MVA) \citep{SonnerupCahill1967} and the
speed from the measured plasma flow, the ambient wind, or a de Hoffmann-Teller (HT) frame determination
\citep{KhrabrovSonnerup1998}. Recent MAVEN observations used exactly this construction to report reconnecting
solar wind current sheets near Mars that are far thicker than the ambient current-sheet population, and
interpreted the excess as reconnection-driven broadening \citep{Zhang2026}.

How accurate is the single-spacecraft construction? Multipoint timing offers the direct test: the crossing delays
$\tau_{pq}$ between spacecraft at separations $\mathbf{r}_{q}-\mathbf{r}_{p}$ determine the normal and boundary
speed without any frozen-in-flow assumption, requiring only local planarity across the baseline,
$(\mathbf{r}_{q}-\mathbf{r}_{p})\cdot\mathbf{m}=\tau_{pq}$ with $\mathbf{m}=\hat{N}/V_{n}$
\citep{Schwartz1998,PaschmannDaly2008,Vogt2011}. Cluster timing established the technique for solar wind
discontinuities at separations of hundreds to thousands of km \citep{Horbury2001}: from four-point timing of
129 discontinuities, \citet{Knetter2004} found MVA normals unreliable and timing normals accurate to about
$10^{\circ}$. At the magnetopause, \citet{Haaland2004} compared four-spacecraft orientation, motion, and
thickness estimates against single-spacecraft determinations. \citet{Mistry2015} used multispacecraft
observations, including tetrahedral timing of exhaust-boundary normals, to constrain the structure and geometry
of confirmed solar wind reconnection exhausts. Two-spacecraft observations have compared the arrival times of
solar wind exhausts \citep{Davis2006}, and widely separated spacecraft have mapped X-line extents
\citep{Phan2006,Lavraud2009}. More recently, \citet{Wang2024} used four-spacecraft timing in a sample of 1,831
solar wind current sheets to benchmark single-spacecraft normal and thickness estimates and the frozen-in
propagation assumption. Their sample was selected as a generic current-sheet population rather than for
reconnection. Reconnection-associated boundaries remain a distinct case because an exhaust boundary can
propagate relative to the surrounding plasma, and a layer hosting locally active reconnection need not be
passively convected. \citet{Wang2023} resolved the electron-scale reconnection region and identified the exhaust
analyzed here as Petschek-like using single-spacecraft analyses of shock normals and speeds, but did not perform
four-spacecraft timing of boundary orientation, motion, or thickness.

The Magnetospheric Multiscale (MMS) mission \citep{Burch2016} provides that opportunity during its solar wind
excursions, at the cost of an unusually small separation: the 2017 tetrahedron spans only $\approx16$ km, about
$0.16\,\di$ and 0.5\%--1\% of the layers of interest. The expected interspacecraft delays are then tens of
milliseconds on magnetic ramps lasting seconds, and we show below that the standard single-threshold
crossing-time method is noise dominated in this regime and can produce large apparent disagreements between
multipoint and single-spacecraft thicknesses. Cross-correlation of the full ramp waveforms, with lag
uncertainties constructed from the sampling cadence, timing closure, and a window jackknife, recovers the
geometry. Here we apply tetrahedral four-spacecraft timing to four current-sheet crossings in the same
solar wind stream: the two boundaries of the identified exhaust and two additional current sheets, one
independently identified as actively reconnecting. We determine their normals and propagation speeds and
compare the resulting thicknesses directly with the standard single-spacecraft estimates for the same crossings.

\section{Event and Data}\label{sec:data}

We use burst-mode MMS data on 2017 November 10: Fluxgate Magnetometer (FGM) fields at 128 samples s$^{-1}$
\citep{Russell2016}, Fast Plasma Investigation (FPI) ion moments at 150 ms \citep{Pollock2016}, and Magnetic
Ephemeris Coordinates (MEC) ephemerides. MMS was at $(9.4,\,21.8,\,6.7)\,R_{\mathrm{E}}$ GSE, upstream of the
bow shock in the pristine solar wind, with mean interspacecraft separation $\approx16$~km
($0.13$--$0.18\,\di$ pairwise). The ambient wind (09:20--09:38 UT averages) had $V_{\mathrm{sw}}=657$ km
s$^{-1}$ along $(-0.996,\,0.084,\,0.031)$, $n_{\mathrm{i}}=4.9$ cm$^{-3}$, $|B|=5.6$ nT, giving
$\di=103$ km and $\vA=55$ km s$^{-1}$. Full-day survey data are away-sector dominated (22 of 24 hourly
median $B_{x}$ values are negative), with no lasting polarity reversal. Across the exhaust the radial field
reverses transiently and returns; the external fields have $B_{x}=-1.6$ and $-3.7$ nT, so the magnetic
record does not show a completed sector-polarity crossing. \citet{Wang2023} use suprathermal-electron
distributions to identify the same excursion as two partial crossings of the heliospheric current sheet. We
therefore leave its global sector-boundary topology open; the timing and thickness results below do not
depend on that classification.

\begin{figure*}
\centering
\includegraphics[width=0.95\textwidth]{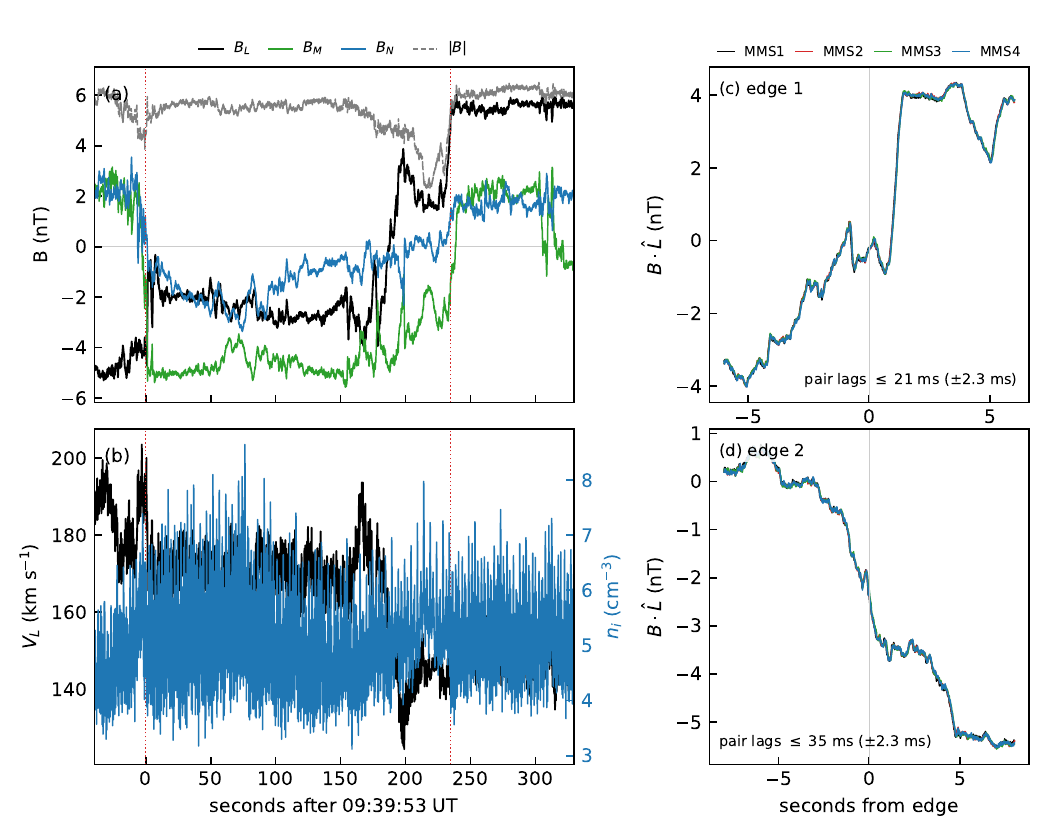}
\caption{The 2017 November 10 reconnection exhaust. (a) MMS1 magnetic field in the exhaust-mean LMN frame;
dotted lines mark the two edges. (b) $V_{L}$ and ion density. (c, d) Four-spacecraft
$B\!\cdot\!\hat{L}$ overlays at the leading and trailing edges; the waveforms are nearly identical,
with pairwise lags of 2--35 ms measured to $\pm2.3$ ms by cross-correlation.\label{fig:overview}}
\end{figure*}

Between 09:39:53 and 09:43:45 UT the spacecraft crossed a reconnection exhaust (Figure \ref{fig:overview});
here and throughout, LMN is the exhaust-mean boundary frame, with $\hat{L}$ along the reconnecting field
component, $\hat{N}$ the boundary normal, and $\hat{M}=\hat{N}\times\hat{L}$ along the guide field. The
reconnecting component reverses from $-4.6$ to $+5.5$ nT across the interval (shears of $97^{\circ}$ and
$73^{\circ}$ at the two edges; guide-field ratio, the mean $B_{M}$ in units of the reconnecting component,
$\approx0.95$), in the range where solar wind reconnection is expected from the observed shear--$\beta$
dependence \citep{Phan2010}. The density is weakly enhanced, and the ion flow inside is deflected along
$-\hat{L}$ by up to 39 km s$^{-1}$, which is $0.81$ of the hybrid Alfv\'en speed
$\vA^{\mathrm{hyb}}=48$ km s$^{-1}$ \citep{CassakShay2007} based on the external reconnecting fields and
densities. The Wal\'en relation \citep{Hudson1970} at the leading edge gives correlation $-0.78$ and slope
$-0.46$; slopes well below unity are typical of solar wind exhausts, whose jets are generally sub-Alfv\'enic
\citep[e.g.,][]{Gosling2012}, so we take this edge as adequately Alfv\'enic for a strong-guide-field exhaust.
The trailing edge is less so ($-0.55$, $-0.18$; Section \ref{sec:results}). The same solar wind stream also
contains the actively reconnecting current sheet at 10:03~UT analyzed by \citet{Wang2023}, which independently
confirms ongoing reconnection in this stream; we analyze it and a further crossing at 10:05~UT below.

\section{Timing Method and Error Budget}\label{sec:method}

At the measured boundary speeds, the interspacecraft delays across the 16 km tetrahedron are of order
20--40 ms, only a few samples at the 7.8 ms magnetometer cadence, while the 10\%--90\% $B_{L}$
transitions span 2.8--7.3~s across the four crossings. Timing a single threshold crossing per spacecraft
then fails: waveform ripple of a few tenths of a nT on a $\sim$1.5 nT s$^{-1}$ ramp displaces individual
level crossings by $\sim$50--200 ms, larger than the interspacecraft delays themselves. A Monte Carlo
with a 58 ms crossing-time jitter (one 7.8 ms sample plus the conservative 50 ms low end of that
displacement range) yields normal cones of $35^{\circ}$--$37^{\circ}$ and speed errors of 14\%--24\%
for every crossing studied here (Figure \ref{fig:methods}), and thickness ratios that scatter between
0.4 and 22 across reasonable threshold and window choices. Any single realization can therefore suggest
a large multipoint--single-spacecraft disagreement that is an artifact of timing noise.

Because the four spacecraft sample essentially identical waveforms at these separations (minimum pairwise
correlation coefficient 0.9998 across the four crossings analyzed), the lags are instead measured by
cross-correlating the full mean-subtracted $B\!\cdot\!\hat{L}$ ramps for all six pairs, with continuous
sub-sample refinement by least-squares waveform alignment. Two empirical diagnostics enter the lag-error
model. The three-spacecraft closure sums $\tau_{pq}+\tau_{qr}-\tau_{pr}$, which vanish identically for a
planar boundary, have rms 1.4 ms (leading edge) and 2.9 ms (trailing edge); a window jackknife
(correlation window rescaled by factors of 0.75--1.1 and shifted by $\pm$15\%) gives consistent per-pair
scatter. Each pair takes
$\sigma_{\tau}=\max(\mathrm{rms}_{\mathrm{closure}}/\sqrt{3},\,\sigma_{\mathrm{jack}},\,
\Delta t/\sqrt{12})$; the 7.8 ms sampling floor $\Delta t/\sqrt{12}=2.26$ ms sets every pair except
two at 10:05 UT, where the jackknife gives 2.7 and 3.3 ms. The MMS timing architecture was designed for 1~ms
timing knowledge across the constellation \citep{Tooley2016}, below the empirical lag uncertainties used here.
Following standard multi-spacecraft timing practice \citep{PaschmannDaly2008,Vogt2011}, the
slowness vector $\mathbf{m}$ is then a weighted least-squares solution over all six pairwise equations,
with $\chi_{\nu}^{2}=0.12$--$0.55$ across the four crossings. Uncertainties are
propagated by Monte Carlo to the normal cone, $V_{n}$, the thickness ratio, and the mutual inclination
of the two edge normals. Two primary limitations affect this error budget. First, closure constraints
test only inconsistencies around the three-spacecraft timing loops; spacecraft-dependent timing offsets
cancel in those loops and remain degenerate with the fitted slowness vector. Second, the
$\chi_{\nu}^{2}$ values of $0.12$--$0.55$ are consistent with conservative lag uncertainties
$\sigma_{\tau}$, but do not establish that the derived error cones are strict upper bounds. A window
sensitivity sweep (27 variants per edge) moves the normals by a median of $0.5^{\circ}$--$1.4^{\circ}$
and the thickness ratios by less than $0.1$, with typical window effects at or below the quoted
statistical errors.

For the single-spacecraft comparison we compute three standard boundary-speed conventions on MMS1: the
local plasma flow projected on the timing normal and averaged across the crossing window,
$|\langle V\rangle\!\cdot\!\hat{N}|$ (the 10\%--90\% transition spans 19--49 ion samples at the
150 ms cadence), which is the construction of \citet{Zhang2026}; the HT frame velocity projected on
the normal, $V_{\mathrm{HT}}\!\cdot\!\hat{N}$ \citep{KhrabrovSonnerup1998}; and the ambient wind
projection $V_{\mathrm{sw}}\!\cdot\!\hat{N}$, with the wind vector measured in the external-side
window of each crossing \citep[the convention of, e.g.,][]{Gosling2005}. The crossing duration
$\delta t$ is the 10\%--90\% transition time of $B_{L}$ between its asymptotic side values, and
thicknesses are $D=$ speed $\times\,\delta t$ for each speed. Throughout, current-sheet thickness
refers to this transition width of an individual boundary; we use exhaust width for the separation
between the two exhaust edges. Because the same $\delta t$ and the same timing normal enter both, the
thickness ratio $\Dfour/\Done$ is identically the speed ratio
$V_{n}/|\langle V\rangle\!\cdot\!\hat{N}|$; the comparison below therefore tests the assumed boundary
speed rather than providing a second, independent spatial measurement. Because a single-spacecraft
observer must also supply the normal, we further determine $\hat{N}$ from MMS1 alone by two standard
routes: unconstrained MVA over the crossing, reporting the intermediate-to-minimum eigenvalue ratio
$\lambda_{2}/\lambda_{3}$ as its conditioning \citep{SonnerupCahill1967}, and the cross-product
$\hat{N}\propto\mathbf{B}_{1}\times\mathbf{B}_{2}$ of the two asymptotic side fields
\citep{GoslingPhan2013}, which is the convention applied at Mars by \citet{Zhang2026}. This tests the
full single-spacecraft recipe, normal as well as speed, against the four-spacecraft solution.

\section{Results}\label{sec:results}

\begin{figure}
\centering
\includegraphics[width=\columnwidth]{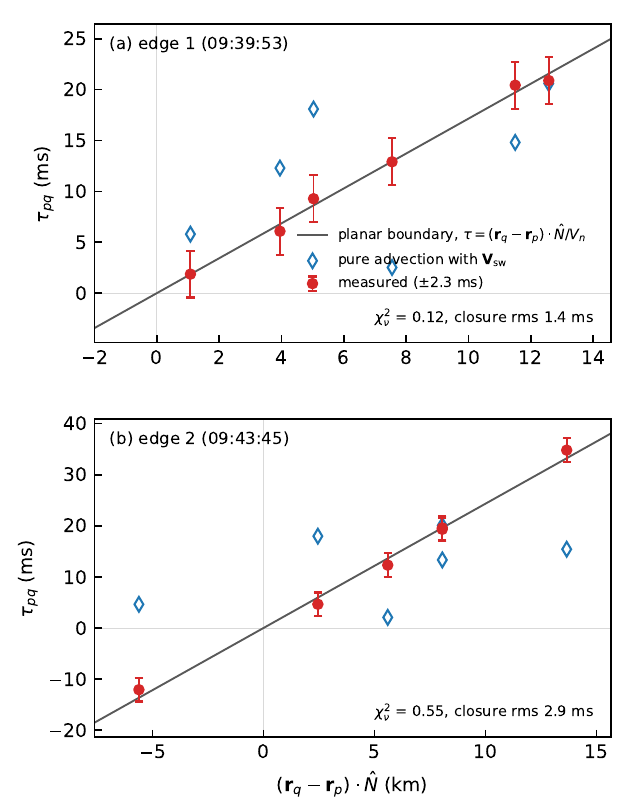}
\caption{Four-spacecraft timing at the leading (a) and trailing (b) exhaust edges. Filled circles show
the measured pairwise lags $\tau_{pq}$ ($\pm2.3$ ms) versus the baseline projection
$(\mathbf{r}_{q}-\mathbf{r}_{p})\!\cdot\!\hat{N}$; the line is the planar-boundary solution
$\tau=(\mathbf{r}_{q}-\mathbf{r}_{p})\!\cdot\!\hat{N}/V_{n}$. Open diamonds show the lags obtained by
projecting each baseline on the external-side solar wind velocity. Their disagreement with the
measurements is primarily geometric, whereas convection with the plasma remains consistent with the
observed lags.\label{fig:timing}}
\end{figure}

Table \ref{tab:edges} collects the measurements. The leading edge (09:39:53 UT) has
$\hat{N}_{1}=(-0.876,-0.472,-0.101)$ GSE with a 68\% normal cone of $5.2^{\circ}$ and
$V_{n}=583\pm41$ km s$^{-1}$; the trailing edge (09:43:45 UT) has
$\hat{N}_{2}=(-0.751,-0.472,-0.462)$ with a $3.6^{\circ}$ cone and $V_{n}=411\pm21$ km s$^{-1}$.
Both boundary velocities are antisunward and comparable to the wind speed, as required for structures
convected by the 657 km s$^{-1}$ flow. The six lags are fitted by a single plane to 0.6 and 1.2 ms,
well inside the 2.3 ms lag uncertainty, and the four waveforms are near-identical
(Section \ref{sec:method}); the spacecraft therefore crossed one coherent surface rather than
uncorrelated fluctuations. The lag pattern is set by the baseline projection on $\hat{N}$ rather than
on the flow: because the normals lie $34^{\circ}$--$46^{\circ}$ from $\mathbf{V}_{\mathrm{sw}}$, the
measured lags depart from
$(\mathbf{r}_{q}-\mathbf{r}_{p})\cdot\mathbf{V}_{\mathrm{sw}}/V_{\mathrm{sw}}^{2}$ by up to a factor
of 5.9 and for one pair in sign, whereas a boundary convecting with the plasma at the measured normal
orientation reproduces them to 1.0 and 2.7 ms (Figure \ref{fig:timing}). With 10\%--90\% transition
times of 5.4 and 7.3 s, the two current layers are $\Dfour=3140$ and 3010 km thick, i.e., $30.5$ and
$29.2\,\di$.

The central result is the single-spacecraft comparison. At the leading edge
$|\langle V\rangle\!\cdot\!\hat{N}|=558$ km s$^{-1}$, $V_{\mathrm{HT}}\!\cdot\!\hat{N}=557$, and
$V_{\mathrm{sw}}\!\cdot\!\hat{N}=550$ km s$^{-1}$: all three conventions agree with each other to
2\% and with the timing speed to 4\%--6\%. The four-spacecraft to single-spacecraft thickness ratio
is $\Dfour/\Done=1.04$, 68\% confidence interval $[0.97,1.12]$ (95\%: $[0.91,1.20]$). At the trailing
edge the three conventions again agree internally ($474$, $473$, $471$ km s$^{-1}$) and the ratio is
$0.87$ $[0.82,0.91]$ (95\%: $[0.79,0.96]$), a small resolved overestimate by the single-spacecraft
method. The boundaries move with the plasma to within
$V_{n}-V_{\mathrm{HT}}\!\cdot\!\hat{N}=+26\pm41$ and $-62\pm21$ km s$^{-1}$ at the leading and
trailing edges, about one Alfv\'en speed. The trailing-edge residual has the sign expected for plasma
entering a slow-mode-like exhaust, but given FPI absolute velocity systematics of a few percent
($\approx25$ km s$^{-1}$) we do not interpret it further. At the exhaust scale, the 232~s crossing
yields widths of $1315\,\di$ and $927\,\di$ along $\hat{N}_{1}$ and $\hat{N}_{2}$, respectively,
based on the multi-spacecraft timing speeds. Single-spacecraft estimates using the local plasma flow
and ambient solar wind velocity agree with these values to within 4\%--15\%.

\begin{figure}
\centering
\includegraphics[width=\columnwidth]{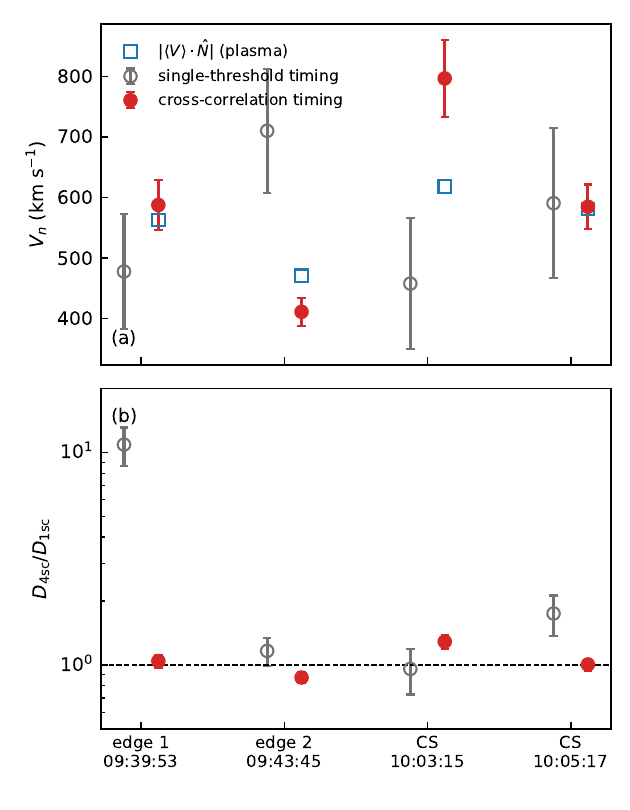}
\caption{Method comparison for the four crossings. (a) Boundary-normal speed from single-threshold
timing (open gray), all-pairs cross-correlation timing (filled red), and the single-spacecraft
$|\langle V\rangle\!\cdot\!\hat{N}|$ estimate (open blue squares). (b) Corresponding thickness ratio
$\Dfour/\Done$; cross-correlation remains near unity except at the actively reconnecting 10:03 UT
sheet, whereas single-threshold timing shows large scatter.\label{fig:methods}}
\end{figure}

Those comparisons hold the timing normal fixed to isolate the speed conventions. A fully
single-spacecraft estimate must also determine $\hat{N}$, the larger error source
(Table \ref{tab:edges}). At the three crossings where the MVA eigenvalue ratio is large
($\lambda_{2}/\lambda_{3}=5$--$13$; the trailing edge and the 10:03 and 10:05 UT crossings), the MVA
normal falls $10^{\circ}$--$16^{\circ}$ from the timing normal. The implied thickness is within a
factor $0.99$--$1.44$ of the four-spacecraft value. At the same three crossings the cross-product
normal performs comparably or better ($6^{\circ}$--$17^{\circ}$). At the high-shear leading edge the
MVA is degenerate ($\lambda_{2}/\lambda_{3}=1.3$), so its minimum- and intermediate-variance directions
interchange, the failure mode documented by \citet{Knetter2004}. The MVA normal is then $42^{\circ}$
off and underestimates the thickness threefold, whereas the cross-product normal, which does not depend
on that ordering, stays within $23^{\circ}$ and a factor $1.5$. For these large-shear crossings, a low
eigenvalue ratio flags the degenerate MVA solution. \citet{Wang2024} report no overall correlation
between MVA accuracy and $\lambda_{2}/\lambda_{3}$, but find a shear-angle dependence: accuracy
improves with increasing $\lambda_{2}/\lambda_{3}$ for $\Delta\theta\gtrsim50^\circ$ and degrades with
increasing $\lambda_{2}/\lambda_{3}$ for $\Delta\theta\lesssim30^\circ$. All four crossings here have
shears of $54^\circ$--$97^\circ$ and follow the large-shear trend, while the eigenvalue ratio is less
reliable for the small-shear sheets that dominate their sample. The dominant uncertainty in a
single-spacecraft thickness is therefore the normal rather than the speed, and the safest
single-spacecraft route, consistent with both our crossings and the statistics of \citet{Wang2024},
is the cross-product normal, with the eigenvalue ratio used only to reject degenerate MVA solutions.

The multipoint timing directly resolves three aspects of the boundary geometry. First, the normals are
quasi-radial, $34^{\circ}$ and $46^{\circ}$ from the ambient flow. Second, the two edges are inclined
to each other by $22^{\circ}$ with 68\% interval $[19^{\circ},26^{\circ}]$
($|\hat{N}_{1}\!\cdot\!\hat{N}_{2}|=0.927$ $[0.899,0.947]$); this is a small but resolved wedge, so
the exhaust is not a plane slab. A single scalar width is only meaningful per normal as quoted above.
Third, the normal magnetic field at the leading edge is nonzero on the external side,
$B_{n}=+1.40$ nT $[1.13,1.64]$, i.e., $0.26|B|$; the trailing edge has $B_{n}=+0.50$ nT
$[0.28,0.73]$ ($0.08|B|$). A resolved finite $B_{n}$ with a sub-unity Wal\'en slope constrains the
local geometry but does not by itself select a unique ideal-MHD discontinuity type. The internal-side
projections are opposite in sign to the external ones ($-1.66$ versus $+1.40$ nT and $-0.19$ versus
$+0.50$ nT). For a single locally planar boundary, $\nabla\!\cdot\!\mathbf{B}=0$ requires continuity
of $B_{n}$ across the surface. The mismatch is therefore consistent with boundary waviness between
the 16 km timing scale and the $10^{4}$ km field-averaging scale, rather than a failure of the timing
solution.

Figure \ref{fig:methods} extends the test to the two additional current sheets in the same stream. At
10:05:17 UT the result repeats: $V_{n}=584\pm37$ versus
$|\langle V\rangle\!\cdot\!\hat{N}|=581$ km s$^{-1}$, ratio $1.01$ $[0.94,1.07]$, with a
$16\,\di$ layer. At the 10:03:15 UT crossing, the actively reconnecting current sheet of
\citet{Wang2023}, the boundary speed exceeds the local plasma speed along the normal:
$V_{n}=797\pm64$ versus $618$ km s$^{-1}$, ratio $1.29$ $[1.18,1.39]$, and the layer is
$33\,\di$ thick. Its external-side $B_{n}=-0.34$ nT $[-0.77,0.11]$ is the only one in the set
consistent with zero (Table \ref{tab:edges}), as expected for a crossing near an actively reconnecting
X line. Our four-spacecraft normal for this sheet agrees to $11^{\circ}$ with the single-spacecraft
$\mathbf{B}_{1}\times\mathbf{B}_{2}$ normal published by \citet{Wang2023}, and a fit of our measured
lags with their normal held fixed has $\chi_{\nu}^{2}=1.4$, so the two independent determinations are
consistent. For comparison, at the leading exhaust edge near 09:40~UT, \citet{Wang2023} adopted the
minimum-variance direction of their 09:39:40--09:40:00 MVA because a magnetic-coplanarity normal lay
within $10^{\circ}$ of it; we reproduce that MVA as degenerate ($\lambda_{2}/\lambda_{3}=1.29$). By
construction, the magnetic-coplanarity and $\mathbf{B}_{1}\times\mathbf{B}_{2}$ formulas yield
orthogonal candidate normals from the same side fields. For those side fields, the timing normal lies
$22^{\circ}$ from the cross-product direction but $89^{\circ}$ from their adopted direction. Because
the MVA at this edge is degenerate ($\lambda_{2}/\lambda_{3}=1.29$), the timing result favors the
cross-product branch over the adopted MVA direction for this crossing. The same figure shows one
realization of the single-threshold estimator applied to all four crossings, giving ratios of 10.9,
1.17, 0.96, and 1.75; the threshold and window sweep in Section \ref{sec:method} shows that such
individual values are not robust.

\begin{deluxetable*}{lcccc}
\tablecaption{Four-spacecraft timing and single-spacecraft comparison for the four current-sheet
crossings.\label{tab:edges}}
\tablehead{\colhead{Quantity} & \colhead{Edge 1 (09:39:53)} & \colhead{Edge 2 (09:43:45)} &
\colhead{CS 10:03:15} & \colhead{CS 10:05:17}}
\startdata
$\hat{N}$ (GSE) & $(-0.88,-0.47,-0.10)$ & $(-0.75,-0.47,-0.46)$ & $(-0.88,+0.48,+0.05)$ &
$(-0.83,+0.52,+0.22)$ \\
68\% normal cone & $5.2^{\circ}$ & $3.6^{\circ}$ & $7.7^{\circ}$ & $6.1^{\circ}$ \\
$\angle(\hat{N}_{\mathrm{MVA}},\hat{N})$; $\lambda_{2}/\lambda_{3}$ & $42^{\circ}$; 1.3 &
$10^{\circ}$; 13 & $12^{\circ}$; 10 & $16^{\circ}$; 5.2 \\
$\angle(\hat{N}_{B_{1}\times B_{2}},\hat{N})$ & $23^{\circ}$ & $7^{\circ}$ & $6^{\circ}$ &
$17^{\circ}$ \\
$V_{n}$ (km s$^{-1}$) & $583\pm41$ & $411\pm21$ & $797\pm64$ & $584\pm37$ \\
$|\langle V\rangle\!\cdot\!\hat{N}|$ (km s$^{-1}$) & $558$ & $474$ & $618$ & $581$ \\
$V_{\mathrm{HT}}\!\cdot\!\hat{N}$ / $V_{\mathrm{sw}}\!\cdot\!\hat{N}$ (km s$^{-1}$) &
$557$ / $550$ & $473$ / $471$ & $620$ / $622$ & $584$ / $578$ \\
$\delta t$ (10\%--90\%, s) & 5.4 & 7.3 & 4.3 & 2.8 \\
$\Dfour$ (km; $\di$) & 3140; 30.5 & 3010; 29.2 & 3410; 33.1 & 1650; 16.0 \\
$\Dfour/\Done$ [68\% CI] & 1.04 [0.97, 1.12] & 0.87 [0.82, 0.91] & 1.29 [1.18, 1.39] &
1.01 [0.94, 1.07] \\
Magnetic shear; guide ratio & $97^{\circ}$; 0.96 & $73^{\circ}$; 0.94 & $76^{\circ}$; 1.27 &
$54^{\circ}$; 1.95 \\
$B_{n}$ external (nT) [68\% CI] & $+1.40$ [1.13, 1.64] & $+0.50$ [0.28, 0.73] &
$-0.34$ [$-0.77$, 0.11] & $+0.79$ [0.42, 1.17] \\
Wal\'en cc; slope & $-0.78$; $-0.46$ & $-0.55$; $-0.18$ & $-0.76$; $-0.23$ &
$-0.64$; $-0.32$ \\
Lag closure rms (ms); $\chi_{\nu}^{2}$ & 1.4; 0.12 & 2.9; 0.55 & 1.4; 0.12 & 2.6; 0.44 \\
\enddata
\tablecomments{Timing uses all-pairs cross-correlation. Lag uncertainties are 2.3 ms except for two
pairs at 10:05:17; quoted uncertainties are 68\% Monte Carlo intervals.
$\Done=|\langle V\rangle\!\cdot\!\hat{N}|\,\delta t$ uses the timing normal $\hat{N}$ to isolate
the speed comparison. $\hat{N}_{\mathrm{MVA}}$ and $\hat{N}_{B_{1}\times B_{2}}$ are the
single-spacecraft MVA and cross-product normals; $\lambda_{2}/\lambda_{3}$ is the MVA
intermediate-to-minimum eigenvalue ratio, with values near unity indicating degeneracy. $\di=103$ km
is the ambient value. The 10:03:15 and 10:05:17 columns use local side windows and refined crossing
epochs of 10:03:14.8 and 10:05:17.0 UT.}
\end{deluxetable*}

\section{Discussion}\label{sec:discussion}

For the reconnection-exhaust boundaries examined here, the local
$|\langle V\rangle\!\cdot\!\hat{N}|$, HT-frame, and ambient-wind speeds agree with the directly
measured boundary speed to 4\%--15\% at the passive exhaust edges, and yield the exhaust width to
within about 15\% along either normal. Including a single-spacecraft normal still recovers the
thickness to within tens of percent, provided the cross-product normal is used or a degenerate MVA
is rejected by its eigenvalue ratio (Section \ref{sec:results}). For the passive exhaust edges
examined here, we find no evidence of a large systematic bias in the single-spacecraft construction
used for the Mars results \citep{Zhang2026}. At the actively reconnecting crossing, however, the
boundary moved 29\% faster than the plasma, so a timing cross-check is warranted where the science
depends on the thickness of a layer hosting ongoing reconnection.

The measured current-sheet thicknesses are large. The four layers span $16$--$33\,\di$ in the
10\%--90\% definition used here; the two exhaust edges and the 10:03 UT sheet are $29$--$33\,\di$,
and the 10:05 UT sheet is $16\,\di$. Solar wind discontinuities and kinetic-scale current-sheet
populations at 1 au and near the Sun are generally characterized by smaller ion-scale widths
\citep{Artemyev2018,Artemyev2019,Vasko2021,Lotekar2022}, although the operational thickness
definitions differ among those studies. The full exhaust width is $927$--$1315\,\di$;
\citet{eriksson2022characteristics} found a 95th-percentile reconnection-exhaust normal width of
$905\,\di$ at 1~au, placing this event in the upper tail of their distribution. These direct
measurements support the interpretation of reconnection-associated broadening, as do the Mars
results at 1.5 au \citep{Zhang2026}.

The two-edge geometry is directly resolved by multipoint timing. The $22^{\circ}$ inclination
(68\% interval $[19^{\circ},26^{\circ}]$) between the two exhaust edges is consistent with the
broader picture that solar wind reconnection exhausts can develop nontrivial boundary structure
\citep{Mistry2015}. With a 16 km baseline, the edge inclination and the resolved external-side
nonzero $B_{n}$ ($0.26|B|$ at the more Alfv\'enic leading edge) provide the kind of geometric
information that Cluster timing obtained at far larger separations \citep{Knetter2004}.
A further methodological result is that, at ion-scale baselines, single-threshold timing is noise 
dominated and can return apparently well-constrained but spurious solutions, whereas waveform cross-correlation 
reduces the normal uncertainties at the exhaust edges from about $\pm36^\circ$ to $\pm4^\circ$--$5^\circ$ without additional data. 
The same problem can arise whenever the physical delays become comparable to the uncertainty in identifying an individual crossing.

Each crossing provides a separate test: the four analyzed here span shears of
$54^{\circ}$--$97^{\circ}$, guide-field ratios of 0.9--2.0, both passive exhaust edges and an
actively reconnecting layer, each with its own timing solution and error budget. The main limitation
is that all four lie in one wind stream; the measurement requires a rare conjunction of burst
telemetry, pristine solar wind, and a compact tetrahedron at an independently confirmed
reconnection site. An automated survey of the surrounding days yielded no further
tetrahedron-quality reconnecting crossings. The trailing edge is only weakly Alfv\'enic; a Wal\'en
slope tests Alfv\'enicity rather than slow-mode character, and we do not assign either boundary a
unique ideal-MHD discontinuity type. Population statistics across wind conditions require future
work: MMS solar wind campaigns at larger separations and archival Cluster tetrahedron intervals
permit the same cross-correlation treatment, as will dedicated multipoint solar wind missions such
as HelioSwarm \citep{Klein2023}.

\section{Conclusions}\label{sec:conclusions}

Using MMS burst data at $\approx16$~km ($0.16\,\di$) separations, we applied four-spacecraft timing to four
current-sheet crossings in a solar wind stream containing a confirmed reconnection exhaust. At the passive
exhaust boundaries, the three standard single-spacecraft speed conventions reproduce the four-spacecraft
thickness to within about 15\%, whereas at the actively reconnecting 10:03~UT layer the timing-based thickness
exceeds the local-plasma estimate by 29\%. Waveform cross-correlation provides millisecond-level timing at this
baseline, where single-threshold timing is dominated by ramp ripple. The exhaust edges are $29$--$31\,\di$
thick, supporting reconnection-associated broadening, while their directly resolved $22^{\circ}$
$[19^{\circ},26^{\circ}]$ mutual inclination establishes a nonparallel two-edge geometry.

\begin{acknowledgments}
We thank J. E. Stawarz for helpful comments on the manuscript. We thank the MMS FGM, FPI, and MEC teams
for the calibrated data, available through the MMS Science Data Center
(\url{https://lasp.colorado.edu/mms/sdc/public/}). We used burst-mode FGM, FPI ion-moment, and MEC data
for 2017 November 10, 09:20--10:10 UT. The analysis scripts are available from the corresponding author
on request.
\end{acknowledgments}

\software{PySPEDAS \citep{Grimes2022}}

\bibliographystyle{aasjournalv7}
\bibliography{references}

\end{document}